\documentclass[%
reprint,
showpacs,preprintnumbers,
 amsmath,amssymb,
 aps,
 pra,
 ]{revtex4-1}
 
\usepackage[T1]{fontenc}
\usepackage{graphicx}
\usepackage{dcolumn}
\usepackage{bm}
\usepackage{geometry}
\usepackage{color}
\begin{document}

\preprint{APS/123-QED}
\title{Two-photon laser excitation of trapped $^{232}$Th$^{+}$ ions via the 402 nm resonance line}
\author{O. A. Herrera-Sancho, M. V. Okhapkin, K. Zimmermann, Chr. Tamm, and E. Peik}

 \email{Corresponding author E-mail address: ekkehard.peik@ptb.de}
 \affiliation{Physikalisch-Technische Bundesanstalt, Bundesallee 100, 38116 Braunschweig, Germany}

\author{A. V. Taichenachev, V. I. Yudin}%

 \affiliation{Institute of Laser Physics, Siberian Branch of RAS, Novosibirsk 630090, Russia}
 \affiliation{Novosibirsk State University, Novosibirsk 630090, Russia}

\author{P. G{\l}owacki}

\affiliation{Pozna\'{n} University of Technology, 60-965 Pozna\'{n}, Poland}

\date{\today}

\begin{abstract}

Experiments on one- and two-photon laser excitation of $^{232}$Th$^+$ ions in a radiofrequency ion trap are reported. As the first excitation step, the strongest resonance line at 402~nm from the (6\emph{d}$^{2}$7\emph{s})\emph{J}=3/2 ground state to the (6\emph{d}7\emph{s}7\emph{p})\emph{J}=5/2 state at 24874~cm$^{-1}$ is driven by radiation from an extended cavity diode laser.
Spontaneous decay of the intermediate state populates a number of low-lying metastable states, thus limiting the excited state population and fluorescence signal obtainable with continuous laser excitation. We study the collisional quenching efficiency of helium, argon, and nitrogen buffer gases, and the effect of repumping laser excitation from the three lowest-lying metastable levels. The experimental results are compared with a four-level rate equation model, that allows us to deduce quenching rates for these buffer gases. Using laser radiation at 399~nm for the second step, we demonstrate two-photon excitation to the state at 49960~cm$^{-1}$, among the highest-lying classified levels of Th$^+$. This is of interest as a test case for the search for higher-lying levels in the range above 55000~cm$^{-1}$ which can resonantly enhance the excitation of the $^{229}$Th$^{+}$ nuclear resonance through an inverse two-photon electronic bridge process.

\end{abstract}


\maketitle

\section{Introduction}\label{sec:intro}

It has been inferred from $\gamma$-spectroscopy of $^{229}$Th that this nucleus possesses an isomeric state 
at an energy of only a few eV above the ground state, establishing the lowest excitation energy that is known in nuclear physics. The most recent experimental value for the $^{229}$Th isomer energy is 7.8(5)~0.5~eV~\cite{Beck:2007, Beck:2009}. This system has stimulated a number of proposals for studies of the effects of atomic electrons on the nuclear transition (see \cite{Matinyan:1998, Tkalya:2003} and references therein). These can be expected to be especially pronounced here because of the matching of the nuclear and electronic excitation energies. The radiative decay of the isomeric state can be strongly enhanced in a so-called electronic bridge process where the nucleus excites the electron shell and a photon is finally emitted in an electronic transition. Vice versa, in an inverse electronic bridge process, laser excitation of the nucleus can be more efficient by making use of an electronic excitation that will resonantly couple to the nuclear moments.

Since the nuclear transition frequency in $^{229}$Th is accessible by frequency upconversion of narrow-bandwidth laser sources and since it is less sensitive to external perturbations than transition frequencies of the electron shell, we have proposed it as the basis of an optical nuclear clock of high precision~\cite{Peik:2003}. Experiments towards this goal are now pursued with different approaches~\cite{Campbell:2011, Rellergert:2010}. Up to now, the large uncertainty of the nuclear transition frequency and its location in the vacuum-ultraviolet range has prevented the observation of excitation or decay of the isomeric state of $^{229}$Th by means of optical spectroscopy. As a technique to achieve resonant excitation of the $^{229}$Th nucleus with the use of conventional laser sources, we have proposed a two-photon inverse electronic bridge process in singly ionized $^{229}$Th~\cite{PorsevPeik:2010}.

The electronic energy level system of Th$^+$ is very complex and is well known only up to an excitation energy of about 7~eV. The classified energy levels belong to the lowest 15 configurations of the three valence electrons. More than 400 energy levels have been identified~\cite{Zalubas:1974} and the wavelengths of about 14000 lines are tabulated~\cite{Palmer:1983}. Recent relativistic Hartree-Fock calculations indicate an exponential increase of the energy level density towards the ionization energy of 11.9~eV~\cite{Dzuba:2010}.

While the high density of electronic energy levels in Th$^+$ increases the probability for a strong resonance enhancement of electronic bridge processes leading to excitation of the $^{229}$Th nucleus, it also poses an experimental problem: Laser excitation of an ensemble of atoms from the ground state to a definite excited state is often inefficient for a multi-level system such as Th$^+$ because spontaneous decay leads to the accumulation of population in metastable levels that are decoupled from the laser. If the number of metastable levels is not too high, additional repumper lasers may be applied for reexcitation from each of those states. Otherwise, atoms in metastable states can be returned to the ground state by inelastic quenching collisions with a buffer gas. The latter approach has already been utilized for laser spectroscopy of Th$^+$ ions: In order to determine nuclear charge radii of the isotopes $^{227}$Th to $^{230}$Th and $^{232}$Th, isotope shifts and hyperfine splittings have been recorded in one- and two-photon laser excitation to levels at 17122~cm$^{-1}$ and 34544~cm$^{-1}$ using Th$^+$ ions in a radiofrequency trap in the presence of helium and hydrogen buffer gases \cite{Kaelber:1989, Kaelber:1992}.

\begin{figure}
\includegraphics[width=0.75\columnwidth]{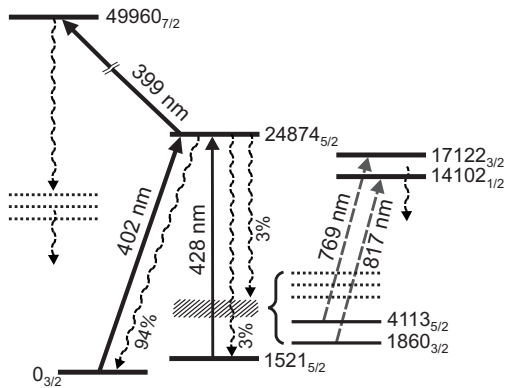}
\caption{\label{fig:excit} Partial level scheme of the thorium ion Th$^+$ showing the main resonance transitions studied in this work and the lowest-lying metastable levels. The hatched area corresponds to a manifold of eleven metastable states which is shown in more detail in the right-hand side of the figure. The level energies are given in cm$^{-1}$ and the subscript denotes the total angular momentum. For the 24874$_{5/2}$ level the spontaneous-decay branching fractions are indicated as given in Ref.~\cite{Nilsson:2002}.}
\end{figure}

Here we present experiments on one- and two-photon laser excitation of trapped Th$^+$ ions, aimed at an investigation of the electronic energy level system in the range around 7-8~eV and towards the excitation of the $^{229}$Th nuclear resonance by an inverse two-photon electronic bridge process. In particular, we investigate excitation of the strongest tabulated line of the Th$^+$ emission spectrum at 402~nm which connects the (6\emph{d}$^{2}$7\emph{s})\emph{J}=3/2 ground state with the (6\emph{d}7\emph{s}7\emph{p})\emph{J}=5/2 state at 24874~cm$^{-1}$~\cite{Palmer:1983}. In the following we label states by their energy in cm$^{-1}$ and their total angular momentum as shown in Fig.~\ref{fig:excit}. Spontaneous decay from the 24874$_{5/2}$ state leads back to the ground state with a rather high probability but a number of low-lying metastable states~\cite{Nilsson:2002} are also populated (see Fig.~\ref{fig:excit}). We measured branching ratios of 11 decay channels after laser excitation in a hollow-cathode discharge. 
With trapped ions, we study the collisional quenching efficiency of helium, argon, and nitrogen buffer gases, and the effect of repumping laser excitation from the three lowest-lying metastable levels. We show that the experimental results are in good agreement with a simple four-level rate equation model. Using laser radiation at 402~nm and at 399~nm, we demonstrate two-photon excitation to the level 49960$_{7/2}$ via the intermediate 24874$_{5/2}$ state. The observation of two-photon excitation to this level in $^{232}$Th$^{+}$ is of interest because it serves as a test case for the search for higher-lying levels which can resonantly enhance the excitation of the $^{229}$Th$^{+}$ nucleus through an electronic bridge process as pointed out above.

\section{Experimental set up}\label{sec:experimental}

\begin{figure}
\includegraphics[width=0.95\columnwidth]{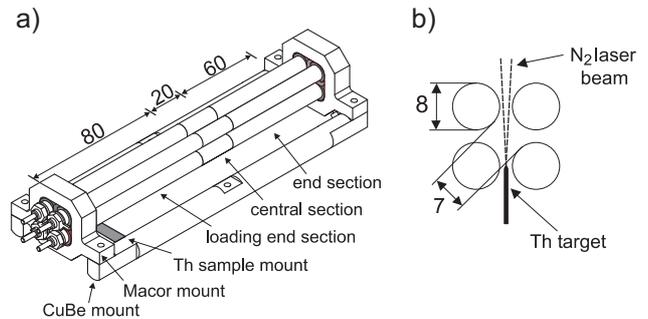}
\caption{\label{fig:trap} Schematic of the linear Paul trap used in the experiments (a) and cross section of the trap electrode arrangement, showing  the position of the Th sample used for ion loading by N$_2$ laser ablation (b). Dimensions are in millimeters.}
\end{figure}

The experiment employs a linear radiofrequency (rf) trap of 160~mm total length (see Fig.~\ref{fig:trap}). The trap electrodes are machined from a CuBe alloy and are divided along their length into three sections which are held at different dc potentials. For ion loading, a metallic $^{232}$Th sample is placed between two electrodes near the end of one of the outer sections. Typically a rf trap drive voltage in the range of 0.5 - 1~kV amplitude at a frequency of 2~MHz is used, corresponding to a Mathieu \textit{q} parameter in the range of 0.2 - 0.4 (for a review see Ref.~\cite{Paul:1990}). The end sections are kept at a dc potential of +40~V relative to the central section for axial confinement.

The trap is mounted in a stainless-steel ultrahigh-vacuum chamber that reaches a base pressure in the range of $1\times10^{-8}$~Pa. For collisional cooling and depopulation of metastable Th$^{+}$ levels by quenching collisions,  either helium, argon or nitrogen at a pressure of up to 0.2~Pa is used as a buffer gas. To produce $^{232}$Th$^{+}$ ions by laser ablation, a nitrogen laser emitting 4~ns pulses with an energy of $\approx100$~$\mu$J at a wavelength of 337~nm is focussed to a spot size of 100~$\mu$m~$\times$~150~$\mu$m on a metallic Th target (see Fig.~\ref{fig:trap}(b)). After 5 to 10 ablation pulses at 0.2~Pa helium buffer gas pressure, more than 10$^{5}$ ions are loaded into the trap (for more details see Ref.~\cite{Zimmermann:2011}). With argon and nitrogen buffer gases the same number of trapped ions is obtained with 2--5 and 1--2 pulses, respectively.  We observe storage times for Th$^+$ in the range of 300--1000~s, limited by the formation of molecular ions in reactions with the background gas~\cite{Kaelber:1992,Chapman:2011}.

For excitation of the 0$_{3/2}$--24874$_{5/2}$ transition (see Fig.~\ref{fig:excit}), a continous-wave (cw) extended-cavity diode laser (ECDL) with a maximum output power of 7~mW at 402~nm and a linewidth of less than 1~MHz is used. The 24874$_{5/2}$--49960$_{7/2}$ transition is excited by a similar laser system emitting at 399~nm. The metastable 1521$_{5/2}$ level is depleted by a frequency-doubled ECDL producing light at a wavelength of 428~nm (see Fig.~\ref{fig:excit}). The output of these lasers is passed to the trap through polarization-maintaining single-mode fibers in order to clean the beam profiles, thus minimizing light scattering off the trap electrodes. Beam diameters in the trap are $\approx$1~mm. For low-resolution spectroscopy over extended wavelength ranges, a tunable modelocked Ti:Sapphire laser system producing pulses of about 3~ps duration and 90~MHz repetition rate is used. Its output wavelength can be tuned in the range of 700--900~nm and in the corresponding second- and third-harmonic ranges.

\begin{figure}
\includegraphics[width=0.9\columnwidth]{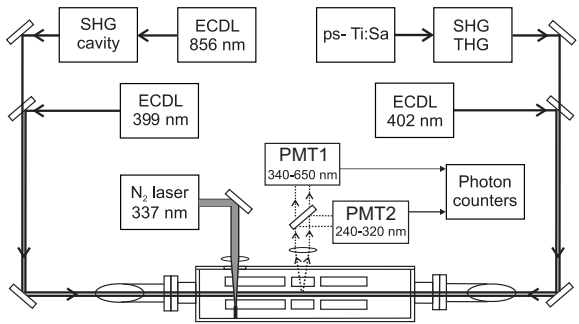}
\caption{\label{fig:optical}Experimental arrangement for laser excitation of trapped Th$^{+}$. Two photomultipliers (PMT) are used for fluorescence detection. A nitrogen laser is used for loading the trap. SHG (THG) second harmonic generation (third harmonic generation). For further details see text.}
\end{figure}

As shown in Fig.~\ref{fig:optical}, the fluorescence emission of the trapped $^{232}$Th$^{+}$ ions is detected with the use of two photomultipliers placed behind a fused-silica collection lens. In order to discriminate between the fluorescence emissions associated with one- and two-step excitations, a combination of spectral filters and a dichroic beamsplitter separates the spectral sensitivity ranges of the photomultipliers. 

\section{Single-wavelength laser excitation of $^{232}$T$\lowercase{\text{h}}$$^{+}$}\label{sec:singlefrequency}

\begin{figure}
\includegraphics[width=1.00\columnwidth]{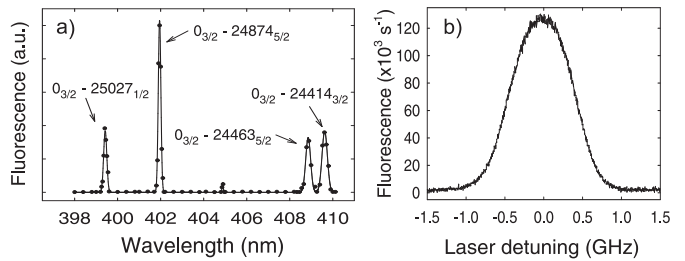}
\caption{\label{fig:2w} Excitation spectra of trapped $^{232}$Th$^{+}$ ions, using (a) the second-harmonic output of a modelocked Ti:Sapphire laser and (b) cw output of an extended-cavity diode laser tuned to the 0$_{3/2}$--24874$_{5/2}$ transition at 402~nm. In (a), the average excitation power is 10~mW and the data points (black dots) are fitted by Lorentzian profiles. In (b), the power is 50~$\mu$W.}
\end{figure}

In an investigation of possible first excitation steps from the ground state,
fluorescence signals  of $^{232}$Th$^{+}$ are readily observed by illuminating the trapped ion cloud with output of the picosecond Ti:Sapphire laser. A typical excitation spectrum of electric-dipole transitions around 400~nm is shown in Fig.~\ref{fig:2w}(a). In order to compensate variations in signal strength caused by fluctuations of the number of loaded ions, the fluorescence signal registered at each laser wavelength was normalized to the signal resulting from resonant cw diode-laser excitation of the 402~nm transition. Helium was used as a buffer gas at a pressure of $\approx$0.2~Pa. The transitions from the ground state displayed in Fig.~\ref{fig:2w}(a) and their relative strengths are in agreement with the data tabulated in Ref.~\cite{Zalubas:1974} for the investigated scan range.

The spectrum resulting from 402~nm cw laser excitation is shown in Fig.~\ref{fig:2w}(b). While the wide linewidths in Fig.~\ref{fig:2w}(a) are determined by the spectral width of the employed laser, the linewidth in Fig.~\ref{fig:2w}(b) is determined by Doppler broadening. For helium pressures above $0.1$~Pa the Doppler width of $\approx 700$~MHz (FWHM) indicates that the trapped ions are collisionally cooled to approximately 300~K for motion along the trap axis. 
With argon and nitrogen buffer gas, cooling to room temperature is achieved at pressures of $\approx 0.2$~Pa and $\approx 0.01$~Pa, respectively. With these gases at $0.2$~Pa pressure, larger fluorescence signals are observed than with helium, because of more efficient collisional quenching of metastable levels (see below). For cw excitation of the 402~nm transition, the strength of the fluorescence signal is strongly affected by population trapping in metastable levels. 

For a quantitative overview on the decay channels of the $24874_{5/2}$ level we performed measurements of the fluorescence emission resulting from laser excitation of Th$^{+}$ ions in a hollow-cathode discharge. The hollow-cathode lamp has a similar construction as described in Refs. \cite{Arcimowicz,Krzykowski}: The cathode is a 2.5 cm long copper cylinder with an inner diameter of 6 mm, covered by a thorium foil on the inner wall, and is cooled by liquid nitrogen. The discharge is operated at a current of 40~mA with argon buffer gas at a pressure between 80~Pa and 100~Pa. A 5~mW laser beam from the 402~nm ECDL, resonant with the 0$_{3/2}$--24874$_{5/2}$ transition, is directed through the cathode. A mirror with a central hole for passage of the laser beam collects the  fluorescence light from the center of the cathode onto the entrance slit of a grating monochromator. 
After passage through the monochromator the light is detected with a photomultiplier and recorded differentially with and without laser excitation, 
in order to separate the laser-induced fluorescence from emission excited by the discharge. The spectral sensitivities of monochromator and photomultiplier are calibrated with a tungsten reference lamp. Since laser stray light would perturb the fluorescence intensity measurement at the wavelength used for excitation, the intensity ratios of the lines at 402~nm, 434~nm and 482~nm were also measured with laser excitation of the  1521$_{5/2}$--24874$_{5/2}$ transition at 428~nm. 
The wide dynamic range of the  detection method allowed us to measure relative intensities of 11 emission lines listed in Ref.~\cite{Zalubas:1974}, complementing a previous study that had reported branching fractions for 6 lines~\cite{Nilsson:2002}. The two weakest lines at 812~nm and at 1073~nm were not observed. Table~\ref{402nmBF} summarizes the results. Based on the observed reproducibility for different discharge conditions, we attribute a relative uncertainty of about 0.1 to the branching fractions for the weak decay channels.  The branching fraction for decay on the resonance line 402~nm back to the ground state of $0.92(3)$ is in good agreement with the value 0.94 from Ref.~\cite{Nilsson:2002} and higher than  the value 0.85 obtained in the analysis in Ref.~\cite{Simonsen:1998}. Cyclic laser excitation of the 402 nm resonance line in a single Th$^{+}$ ion can therefore be expected to result in about 15 fluorescence photons before decay into a metastable level occurs.

\begin{table}[ht]
\caption{ Branching fractions (BF) for decay of the Th$^+$ level 24874$_{5/2}$ to various lower-lying metastable states.}
\label{402nmBF}
\begin{tabular}{cccc}
\hline
\hline
lower level  &  $\lambda_{air}$ & 	BF 			 & BF  \\
(cm$^{-1}$)  &    (nm)					&(this work) &(Ref. \cite{Nilsson:2002})\\
\hline
 &  &  &     \\ 
0& 401.9 & 9.2$\cdot$10$^{-1}$       & 9.4$\cdot$10$^{-1}$ \\ 
1521 & 428.1 & 3.3$\cdot$10$^{-2}$  & 3$\cdot$10$^{-2}$  \\ 
1860 & 434.4 & 2.7$\cdot$10$^{-2}$  & 1$\cdot$10$^{-2}$ \\ 
4013 & 481.5 & 3$\cdot$10$^{-5}$    & - \\ 
4146 & 482.3 & 7$\cdot$10$^{-3}$    & 1$\cdot$10$^{-2}$ \\ 
7001 & 559.4 & 6$\cdot$10$^{-3}$    & 1$\cdot$10$^{-2}$ \\ 
8460 & 609.1 & 5$\cdot$10$^{-4}$    & - \\ 
8605 & 614.5 & 2$\cdot$10$^{-3}$    & 4$\cdot$10$^{-3}$ \\ 
9061 & 632.2 & 2$\cdot$10$^{-4}$    & - \\ 
9711 & 659.4 & 4$\cdot$10$^{-4}$    & - \\ 
13250 & 860.1 & 4$\cdot$10$^{-5}$   & - \\ 
\hline
\end{tabular}
\end{table} 

\section{Effective 4-level model for collisional quenching and repumping}\label{sec:theory}

\begin{figure}
\includegraphics[width=0.60\columnwidth]{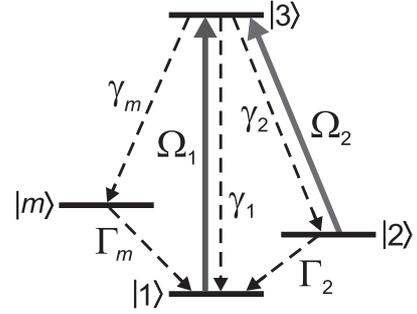}
\caption{\label{fig:4system} Energy level system considered in the rate equation model of Eq.~\ref{rate}. The levels $\left|1\right\rangle$, $\left|3\right\rangle$, $\left|2\right\rangle$, and $\left|m\right\rangle$ represent the ground state, the 24874$_{5/2}$ and 1521$_{5/2}$ levels, and a manifold of eleven metastable states, respectively  (cf. Fig.~\ref{fig:excit}).}
\end{figure}

In order to obtain a simple analytic description, we approximate the Th$^{+}$ level system up to the 24874$_{5/2}$ level by the four-level system shown in Fig.~\ref{fig:4system}: the 402~nm transition from the ground state $\left|1\right\rangle$ to state $\left|3\right\rangle$ is driven with Rabi frequency $\Omega_{1}$, the metastable state $\left|2\right\rangle$ is depleted by 428~nm repumping light with Rabi frequency $\Omega_{2}$, and the manifold of eleven higher-lying metastable states is represented by a single level $\left|m\right\rangle$. Here we assume that $\left|m\right\rangle$ is depleted only by quenching collisions. The excited state $\left|3\right\rangle$ radiatively decays into the states $\left|1\right\rangle$, $\left|2\right\rangle$ and $\left|m\right\rangle$ with rates $\gamma_{1}$, $\gamma_{2}$ and $\gamma_{m}$. Their sum, $\gamma$=$\gamma_{1}+\gamma_{2}+\gamma_{m}$, determines the radiative lifetime of the state $\left|3\right\rangle$ as $\tau=\frac{1}{\gamma}$. For our analysis we use the experimental lifetime value $\tau  =23~\text{ns}$~\cite{Simonsen:1998}. We assume that $\gamma_{1}$=$b\gamma$ and $\gamma_{2}$=$\gamma_{m}$=$\left(1-b\right)\frac{\gamma}{2}$, and $b=0.94$ as the branching fraction for decay to the state $\left|1\right\rangle$, because the branching fractions for decay to the 1521$_{5/2}$ level and for decay to $\left|m\right\rangle$ are approximately equal (see Ref. \cite{Nilsson:2002} and Table~\ref{402nmBF}). The populations  of $\left|2\right\rangle$ and $\left|m\right\rangle$ can decay to the ground state through quenching collisions with buffer gas and the corresponding rates are denoted by $\Gamma_{2}$ and $\Gamma_{m}$. Neglecting the light-induced coherence between states $\left|1\right\rangle$ and $\left|2\right\rangle$, the population distribution among the levels can be described by rate equations for the population probabilities $p_i$ ($i=1,2,3,m$). In the steady-state limit the rate equations of the four-level system shown in Fig.~\ref{fig:4system} can be expressed as:
\begin{align} \label{rate}
-\gamma S_1p_1 + \Gamma_2p_2 + \gamma_1p_3 + \gamma S_1 p_3 + \Gamma_mp_m &= 0,\nonumber\\
-\gamma S_2p_2 - \Gamma_2p_2 + \gamma_2p_3 + \gamma S_2 p_3 &= 0,\nonumber\\
\gamma S_1p_1 + \gamma S_2p_2 - \gamma p_3 - (S_1 + S_2)\gamma p_3 &= 0,\nonumber\\
\gamma_mp_3 - \Gamma_mp_m &= 0,
\end{align}
with the normalization condition $\sum p_i = 1$. Here $S_j$ ($j=1,2$) are the saturation parameters for the transitions $\left|j\right\rangle-\left|3\right\rangle$. Assuming thermalization of velocities due to both velocity-changing collisions and interaction with the trap potential, and vanishing detunings from the transitions $\left|j\right\rangle-\left|3\right\rangle$, we approximate the optical pumping rates as:
\begin{equation} \label{maxwellization}
\gamma S_j \approx
2\sqrt{\pi}\frac{\Omega^{'2}_{j}}{kv}\;,
\end{equation}
where $kv \approx 2\pi\times 360~$~MHz is the Doppler width.  For linearly polarized laser fields, the effective Rabi frequencies $\Omega^{'}_{j}$ are obtained by averaging over the Zeeman sublevels:
\begin{equation} \label{zeeman}
\Omega^{'2}_{j}=\frac{(2J_{3}+1)}{3(2J_{j}+1)}\Omega^{2}_{j},
\end{equation}
where $J_{j}$ is the total angular momentum of the level $\left|j\right\rangle$.

Solving the rate equations, the population of state $\left|3\right\rangle$ is:
\begin{eqnarray}\label{normalization}
&{}p_3=[2G_2G_mS_1 + 2G_mS_1S_2]\\
&{}\times\{2G_2G_m + [(1-b)(G_2 + G_m)+4G_2G_m]S_1+\nonumber\\
&{}+[(1+b)G_m+2G_2G_m]S_2+ [(1-b)+6G_m]S_1S_2\}^{-1}\nonumber\\\nonumber
\end{eqnarray}
where $G_k=\Gamma_k/\gamma$, ($k=2,m$). The observed fluorescence rate is proportional to $p_3$. If both optical excitations $\left|j\right\rangle-\left|3\right\rangle$ saturate the corresponding transitions, $S_1 \gg (1+b)G_2/(1-b)$ and $S_2 \gg (G_2+G_m)$, the fluorescence rate is limited by the quenching relaxation of the state $\left|m\right\rangle$:
\begin{equation} \label{repumpingon}
p_3 \approx \frac{2G_m}{1-b}\;.
\end{equation}
In the absence of the repumping field ($S_2=0$), level $\left|2\right\rangle$ is depleted only by quenching collisions. In this case the fluorescence rate is reduced and can be expressed as follows:
\begin{equation} \label{repumpingoff}
p_3 \approx \frac{2G_m}{1-b}\frac{G_2}{G_2+G_m}\;.
\end{equation}
According to Eq.~\ref{repumpingon} and Eq.~\ref{repumpingoff}, the fluorescence enhancement factor $\textit{g}$ due to repumping from level $\left|2\right\rangle$ obtained in the limit of high 402~nm and 428~nm laser powers is given by $\textit{g}$=1+$\Gamma_{m}$/$\Gamma_{2}$. 

\section{Quenching rates and population of the 24874$_{5/2}$ level}\label{sec:population}

Based on the results of the four-level model description and using repumping excitation from the 1521$_{5/2}$ level, we determine the collisional quenching rates of this level and of the manifold of the other metastable levels which are populated during continuous excitation of the 402~nm 0$_{3/2}$--24874$_{5/2}$ transition. Quenching rate coefficients are determined for helium, argon, and nitrogen buffer gases. In order to find conditions which minimize the population of metastable levels and thus maximize the population of the 24874$_{5/2}$ state, we also investigate the effect of additional repumping excitation from the manifold of the levels above the 1521$_{5/2}$ level, which are described as the effective level $\left|m\right\rangle$ in the model calculation (see Fig.~\ref{fig:4system}).

\begin{figure}
\includegraphics[width=0.95\columnwidth]{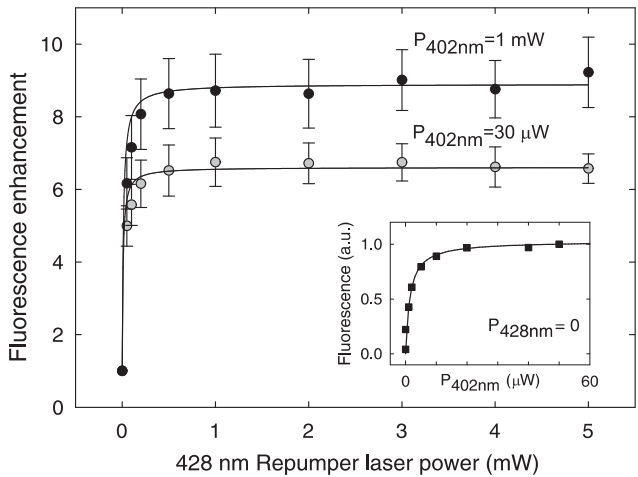}
\caption{\label{fig:saturation402_repumper428} Fluorescence signal of trapped Th$^{+}$ ions resulting from resonant laser excitation of the 402~nm 0$_{3/2}$--24874$_{5/2}$ transition, showing the fluorescence enhancement due to repumping excitation at 428~nm for two settings of 402~nm excitation power (Full and open circles) with helium buffer gas. The data points are normalized to the fluorescence levels observed without 428~nm excitation. The lines correspond to solutions of Eq.~\ref{normalization} using excitation and quenching rate parameters corresponding to the conditions of the experiment (see text). The data points in the inset show the variation of the fluorescence signal with 402~nm laser power observed in the absence of 428~nm excitation. The solid line corresponds to the solution of Eq.~\ref{normalization} using the same parameters as in the main figure.}
\end{figure}

The data points in Fig.~\ref{fig:saturation402_repumper428}(a) show the relative increase  in the fluorescence signal at 402~nm which results from repumping excitation of the 1521$_{5/2}$--24874$_{5/2}$ transition at 428~nm. Helium was used as a buffer gas at 0.2~Pa pressure. The data were obtained for two settings of 402~nm laser power which differ by more than one order of magnitude. It appears that the fluorescence signal is enhanced by up to a factor of nine at a repumping laser power above 0.5~mW. The enhancement is less pronounced if the 402~nm power is reduced. If argon is used at the same pressure, the repumping exitation at 428~nm leads to a maximum fluorescence enhancement $g\approx 2$, and using nitrogen, only a negligible fluorescence enhancement was observed. At variance with helium, with argon and nitrogen buffer gases a weak fluorescence signal is also observed if the $^{232}$Th$^{+}$ ions are illuminated only by the 428~nm light without excitation at 402~nm. This observation indicates that the 1521$_{5/2}$ level can be populated by Th$^{+}$--Ar and Th$^{+}$--N$_2$ collisions. 

Using the experimental data for high and low 402~nm excitation power shown in Fig.~\ref{fig:saturation402_repumper428} together with the calculated optical excitation rates and the natural lifetime and spontaneous-decay branching fractions of the 24874$_{5/2}$ state from Ref. \cite{Nilsson:2002} (see Table~\ref{402nmBF}), we find a good agreement between the observed fluorescence enhancement factor \textit{g} and the predictions of Eq.~\ref{normalization} with $\Gamma_{2}$~=~20~s$^{-1}$ for helium. As shown in Table~\ref{gases}, we find in a similar way $\Gamma_{2}$~=~1.5$\cdot$10$^{3}$~s$^{-1}$ for argon, and $\Gamma_{2}$~=~5.0$\cdot$10$^{4}$~s$^{-1}$ for nitrogen, at 0.2~Pa pressure for all gases. Using the branching fraction $b=0.92$  measured in our experiments (see Table~\ref{402nmBF}), we obtain slightly higher quenching rates (for example, for helium, $\Gamma_{2}$~=~25~s$^{-1}$).
The diatomic buffer gas N$_2$ yields much higher quenching efficiency than the noble gases. A similar behaviour is known for the collisional quenching rates of the metastable $D$ levels in alkali-like ions~\cite{Madej:1990,Knoop:1998}. 

The inset of Fig.~\ref{fig:saturation402_repumper428} shows that Eq.~\ref{normalization} also accurately describes the observed saturation behaviour of the 402~nm excitation without 428~nm repumping. In this case, population of the 1521$_{5/2}$ state reduces the effective saturation power for excitation of the 402~nm transition to a few microwatts corresponding to an intensity of $ \approx 0.3$~mW/cm$^2$, i.e. two orders of magnitude below the saturation intensity for the two-level system 0$_{3/2}$--24874$_{5/2}$. With helium buffer gas, the observed variation of $\textit{g}$ with pressure points to a linear pressure dependence of the rate $\Gamma_{2}$ and to a less than proportional variation of $\Gamma_{m}$ with pressure. At a lower helium pressure of 0.02~Pa, the 428~nm repumping excitation enhances the fluorescence signal by $\textit{g}$~$\sim$~20.

\begin{table}[b]
\caption{\label{gases}Quenching rates $\Gamma_2$ for different buffer gases at 0.2~Pa pressure and maximum fraction of Th$^{+}$ ions excited to the 24874$_{5/2}$ level, and estimated population $p_{3}$ in the state $\left|3\right\rangle$ without and with repumping at 428~nm.}
\begin{ruledtabular}
\begin{tabular}{cccc}
Buffer gas & $\Gamma_2$        &$p_{3}$          &$p_{3}$ \\
					 &(s$^{-1})$				 &(without 428 nm) & (with 428~nm)\\
\hline
helium     &20                &1.5$\cdot$10$^{-5}$&1.2$\cdot$10$^{-4}$\\
argon      &1.5$\cdot$10$^{3}$&5.0$\cdot$10$^{-4}$&1.0$\cdot$10$^{-3}$\\
nitrogen   &5.0$\cdot$10$^{4}$ &5.0$\cdot$10$^{-3}$&                    \\
\end{tabular}
\end{ruledtabular}
\end{table}

Using time-separated pulsed 402~nm and 428~mn excitation we obtain a direct measurement of the quenching rate $\Gamma_{2}$ of the 1521$_{5/2}$ level. 
The 1521$_{5/2}$ level was populated via the 24874$_{5/2}$ state during a 2~ms pulse of the ECDL at 402~nm. After a variable time delay, the remaining population of the 1521$_{5/2}$ level is determined by applying  a pulse from the 428~nm ECDL and recording the time-integrated fluorescence signal at 402~nm. A similar method was used in Refs.~\cite{Madej:1990, Knoop:1998}. We observe an exponential decay of the fluorescence signal as a function of the time delay between the 402~nm and 428~nm excitation and a linear dependence of the decay rate on He pressure of 63(3)~s$^{-1}$Pa$^{-1}$. For a pressure of 0.2~Pa, this results in $\Gamma_{2}$=13(2)~s$^{-1}$, in agreement with the value derived above with Eq.~\ref{normalization} from the data for continous excitation. 

In order to maximize the efficiency of laser excitation to the 24874$_{5/2}$ level, we investigated extended repumping schemes where in addition to the 1521$_{5/2}$ level also higher-lying metastable levels are depleted by laser excitation (see Fig.~\ref{fig:excit}). The output wavelength of the pulsed Ti:Sapphire laser (see Sec.~\ref{sec:experimental}) was tuned to suitable repumping transitions for the six lowest metastable levels above the 1521$_{5/2}$ level that are populated by spontaneous decay from the 24874$_{5/2}$ level. For helium buffer gas at 0.2~Pa, a fluorescence enhancement of approximately a factor of three is observed for excitation of the 4113$_{5/2}$--17122$_{3/2}$ transition and for a number of other transitions originating from the 4113$_{5/2}$ level. Excitation from higher-lying metastable levels does not lead to any significant increase in the fluorescence signal. The depletion of the 1860$_{3/2}$ level increases the fluorescence signal only if there is no repumping excitation from the 1521$_{5/2}$ level. This can be explained by a collision-induced population transfer between the 1521$_{5/2}$ and 1860$_{3/2}$ levels whose energy difference is comparable to the kinetic energy of the buffer gas atoms. With argon or nitrogen buffer gas, repumping from the energy levels above the 1521$_{5/2}$ level does not result in any significant fluorescence enhancement.

Table~\ref{gases} shows the inferred effective quenching rates together with the calculated population probability of the 24874$_{5/2}$ level under cw laser excitation for the three gases used in our investigations. For helium, additional repumping from the 4113$_{5/2}$ level yields a maximum population of 3.6$\cdot$10$^{-4}$ in the 24874$_{5/2}$ level. 
For all experimental conditions studied here, population trapping in low-lying metastable levels limits the obtainable population of the 24874$_{5/2}$ state. In previous experiments with trapped Th$^+$ ions, molecular hydrogen and helium buffer gas were compared and hydrogen was found to produce $\approx 100$ times higher fluorescence signal at the same pressure~\cite{Kaelber:1992}. In our experiment, nitrogen shows an even stronger quenching efficiency.   
From the maximum 402~nm fluorescence signal obtained with nitrogen we deduce that more than 100~photons/s per ion are detected, providing efficient diagnostics of the trapped ions on the timescale expected for the decay of the $^{229}$Th isomeric state~\cite{Porsev:2010}. While nitrogen produces the highest fluorescence rate and most efficient collisional cooling, the noble gases offer the advantage that they can be purified more efficiently using getter materials and cryogenic traps and therefore permit longer storage times for Th$^+$.
     
With the use of pulsed optical excitation, even without repumping excitation a significantly higher population in the 24874$_{5/2}$ level can be achieved over times short compared to $1/\Gamma_2$ and $1/\Gamma_m$. This could be demonstrated by using an acousto-optical modulator (AOM) to switch on the 402~nm excitation repeatedly with a rise time of approximately 1~$\mu$s after dark periods long compared to $1/\Gamma_2$. The resulting initial fluorescence intensity is more than 30 times larger than the steady-state value obtained with 402~nm cw excitation and helium buffer gas. The decay of the fluorescence signal to the level corresponding to cw excitation was dominated by two time constants in the range of a few 100~ns and 1~ms. The longer time constant might reflect the collision-induced population transfer between the various metastable levels (see above). 
The observed fast time constant can not be directly associated with the spontaneous decay rates to metastable levels because also transit-time and saturation effects and the rise time of the AOM shutter are expected to strongly affect the temporal variation of the fluorescence signal on this time scale. 

\section{Two-photon laser excitation of trapped $^{232}$T\lowercase{h}$^{+}$ ions}\label{sec:two photon excitation}

The level 49960$_{7/2}$ shown in Fig.~\ref{fig:excit} is one of the highest-lying tabulated energy levels that can be excited by an electric dipole transition from the 24874$_{5/2}$ state. The radiative lifetime of this level appears to be unknown. Apart from the 399~nm transition to the 24874$_{5/2}$ level, there are 17  tabulated decay channels to other levels~\cite{Zalubas:1974}. The energy of the 49960$_{7/2}$ state corresponds to 6.2~eV which is close to the range of the expected isomeric excitation energy of $^{229}$Th$^{+}$.

\begin{figure}
\includegraphics[width=1.0\columnwidth]{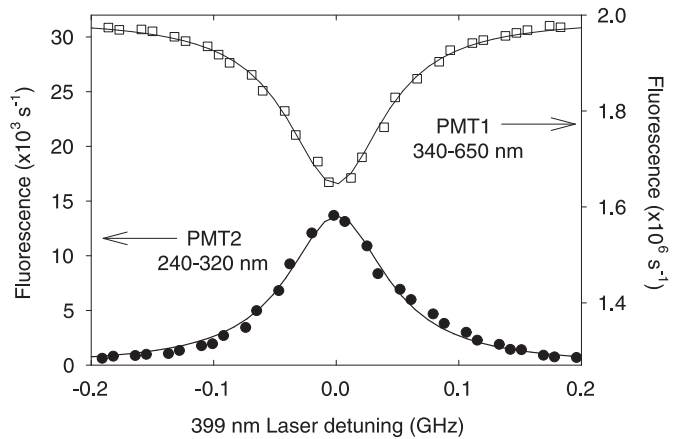}
\caption{\label{fig:twophotonsingle} Two-photon laser excitation spectrum for counter-propagating laser beams, showing fluorescence signals from the upper 49960$_{7/2}$ level (filled circles) and from the intermediate 24874$_{7/2}$ state (open squares). Radiation at 402~nm is tuned to the line center of the 0$_{3/2}$--24874$_{5/2}$ transition and 399~nm light is scanned in the range of the 24874$_{5/2}$--49960$_{7/2}$ transition. The solid lines are  fits to a Lorentzian line shape.}
\end{figure}

In our experiment, trapped Th$^{+}$ ions collisionally cooled by buffer gases at 0.2~Pa pressure are continuously excited by 402~nm radiation tuned to the line center of the 0$_{3/2}$--24874$_{5/2}$ transition and by 428~nm repumping light. A beam of 399~nm light from an ECDL whose frequency is scanned across the 24874$_{5/2}$--49960$_{7/2}$ transition is overlapped co- or counterpropagating with the 402~nm beam. Excitation to the 49960$_{7/2}$ level is detected by monitoring the fluorescence emission in the wavelength range below 320~nm. Figure~\ref{fig:twophotonsingle} shows the two-photon excitation spectrum observed for counterpropagating 402~nm and 399~nm excitation beams using helium buffer gas at 0.2~Pa. The fluorescence signal observed in the range 340--650~nm, also shown Fig.~\ref{fig:twophotonsingle}, indicates the population of the intermediate 24874$_{5/2}$ level. The resonant reduction of this fluorescence shows that a substantial fraction of ions ($\approx$20~\% for counterpropagating beams) is transferred from the 0$_{3/2}$--24874$_{5/2}$ excitation cycle if the 24874$_{5/2}$--49960$_{7/2}$ transition is resonantly excited. This shows that also in the case of the complex level-structure of Th$^+$ efficient two-photon excitation of highly excited states can be achieved. 

\begin{figure}
\includegraphics[width=1.00\columnwidth]{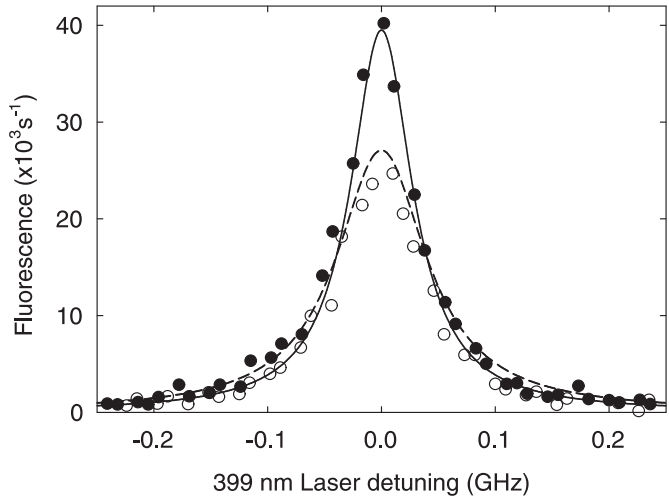}
\caption{\label{fig:twophotonfit} Fluorescence signal of trapped $^{232}$Th$^{+}$ ions resulting from two-photon laser excitation of the 49960$_{7/2}$ level. The spectra are detected by monitoring the fluorescence emission in the wavelength range 240--320~nm. Nitrogen at 0.2 Pa is used as a buffer gas. Filled circles: counterpropagating beams; open circles: copropagating beams. The black (counterpropagating beams) and dashed (copropagating beams) lines correspond to calculations based on the model from Ref.~\cite{Bjorkholm:1976}.}
\end{figure}

Generally, one expects that both, stepwise and direct two-photon excitation processes contribute to the population of the 49960$_{7/2}$ level. 
With counterpropagating beams, the linewidth of the two-photon resonance is narrower and the fluorescence signal is approximately 1.5 times higher than with copropagating beams (see Figure~\ref{fig:twophotonfit}). In the copropagating case the sub-Doppler resonance populating the 49960$_{7/2}$ level is formed predominantly by stepwise two-photon excitation, while in the counterpropagating case an additional contribution of direct two-photon excitation appears. A similar feature was observed in a previous investigation on two-photon excitation between lower-lying energy levels of trapped Th$^{+}$ ions~\cite{Kaelber:1992}. 

In Figure~\ref{fig:twophotonfit} the experimental data for co- and counterpropagating beams are plotted together with  calculated lineshapes based on the model presented in  Ref.~\cite{Bjorkholm:1976}. The homogeneous linewidth of the 402 nm transition was inferred from Doppler-free saturation resonances of the transition at 402~nm. Using nitrogen buffer gas, we here observe a minimum width of 26(2)~MHz (FWHM), significantly higher than the natural linewidth of the 24874$_{5/2}$ state of 7~MHz. Apart from smaller contributions from saturation, collisions and transit effects, the dominant broadening mechanism can likely be attributed to a frequency modulation resulting from the driven micromotion of the ions in the trap. For the fitting of the two-photon lineshapes, good agreement with theory was obtained for a width of the 24874$_{5/2}$--49960$_{7/2}$ resonance which is four times bigger than the width of the 24874$_{5/2}$ resonance. The resulting linewidths for excitation with co- and counterpropagating laser beams are 95~MHz and 66~MHz, respectively. 

Using the pulsed picosecond Ti:Sapphire laser for the second excitation step results in weak fluorescence signals which did not allow us to identify the excitation of higher-lying states. The broad spectrum emitted by this laser excites several resonances from low-lying metastable levels simultaneously.
\section{Conclusion}
In conclusion, we have demonstrated two-photon excitation of Th$^+$ through the intermediate state at 24874$_{5/2}$. In order to maximize the  population of excited states we investigated collisional quenching of metastable states with different buffer gases and repumping with additional lasers. In continous excitation with two diode lasers, we have shown the efficient two-photon excitation of a highly excited state 
in this complex level-structure. 
This sets the ground for a comprehensive investigation of the electronic level structure of Th$^+$ in the energy range of the  $^{229}$Th isomeric state \cite{Porsev:2010} and for the search for a resonant two-photon electronic bridge excitation of the $^{229}$Th nucleus \cite{PorsevPeik:2010} over the wide present uncertainty range for the transition energy.   
For this we plan to use the third harmonic of a pulsed Ti:Sapphire laser with a linewidth in the GHz range in combination with synchronized 402~nm ECDL pulses. A comparison of the  fluorescence signals observed with $^{229}$Th$^+$ and with $^{232}$Th$^+$ shall permit the unambiguous identification of the signature of the unique nuclear structure of $^{229}$Th.   
\begin{acknowledgements}
We would like to thank Th. Leder for his expert technical support in the construction of the experiment. This work was
supported by DFG through QUEST. OAHS acknowledges support from DAAD (Grant No. A/08/94804), ITCR (Grant No. 83-2008-D), and MICIT (Grant No. CONICIT 079-2010). AVT and VIY are supported by  DFG/RFBR (Grant No. 10-02-
91335), RFBR (Grants No. 10-02-00406, No. 11-02-00775, and No. 11-02-01240), RAS, Presidium SB RAS, and the federal programs ``Development of scientific potential of higher school 2009--2010'' and ``Scientific and pedagogic personnel of innovative Russia 2009--2013''. The work of PG was funded by the COST action MP1001 IOTA.
\end{acknowledgements}

\end{document}